

\def\ifundefined#1{\expandafter\ifx\csname
#1\endcsname\relax}

\newcount\eqnumber \eqnumber=0
\def\beq{ \global\advance\eqnumber by 1 $$ }
\def\eeq{ \eqno(\the\eqnumber)$$ }
\def\label#1{\ifundefined{#1}
\expandafter\xdef\csname #1\endcsname{\the\eqnumber}
\else\message{label #1 already in use}\fi}
\def\(#1){(\csname #1\endcsname)}
\def\puteqno{\global\advance \eqnumber by 1 (\the\eqnumber)}

\newcount\refno \refno=0
\def\[#1]{\ifundefined{#1}\advance\refno by 1
\expandafter\xdef\csname #1\endcsname{\the\refno}
\fi[\csname #1\endcsname]}
\def\refis[#1]{\item{\csname #1\endcsname.}}


\baselineskip=18pt
\magnification=1200
{\it   }
\vskip 6cm
\centerline {\bf TWO DIMENSIONAL BARYONS IN }
\centerline{\bf THE LARGE N LIMIT}
\vskip 2cm
\centerline{P.F.Bedaque, I.Horvath and S.G.Rajeev}
\vskip 1cm
\centerline{\it{Department of Physics and Astronomy}}
\centerline{\it{University of Rochester}}
\centerline{\it{Rochester, N.Y. 14627}}
\vskip 2cm
\centerline{\bf{Abstract}}
\vskip 1cm
We propose a bilocal field theory for mesons in two dimensions
obtained as a kind of non local bosonization
of two dimensional QCD. Its semi-classical expansion is equivalent to the
$1/N_c$ expansion of QCD. Using an ansatz we reduce the classical
equation of motion of this theory in the baryon
number one sector to a relativistic Hartree equation and solve it numerically.
This (non topological) soliton is identified with the baryon.
\vfill
\eject
\bigskip
Its widely believed that quantum chromodynamics (QCD) is the theory of
strong interactions. However the degrees of freedom of its original
lagrangian (quarks and gluons) do not correspond to the physical particles
(hadrons). Besides, many properties of the physical particles have an
intrinsically non-perturbative character, making difficult to make predictions
starting from first principles. One possible strategy is to construct
effective theories valid in the low energy domain.
In four dimensions the spontaneous breakdown of chiral symmetry leads to
the non-linear sigma model that describes the
lightest mesons (pions), that appear
in this context as Goldstone bosons. An interesting fact about this model
is the presence of soliton solutions
(the so called Skyrme term [1] has to be included to stabilize
these solutions). After the inclusion of the Wess-Zumino
term [2], this soliton is a fermion, has baryon number one, and consequently,
can be identified with the nucleon. In this way,
although nucleons are not explicit
in the lagrangian, their properties can be studied, with reasonable agreement
with experiment [3]. Arguments using the $1/N_c$ expansion [4]also suggest that
nucleons are solitons [5].
   In some aspects two dimensional QCD resembles the four dimensional case
(for example, it is confining) but it is much simpler, providing a
place to test ideas too difficult to apply directly in four dimensions, and
it has been numerically solved using a discrete light cone quantization [6].
One can take an attitude similar to
the four dimensional case and consider low energy effective theories [7], even
though chiral symmetry is not spontaneously broken. But in two dimensions
one can do better: to write an equivalent lagrangian involving
only meson fields without relying on low energy approximations. The price
to pay for this is that the mesons fields are not local. To see this let us
start with the QCD action with one flavor
\beq S=\int d^2x \sqrt{-g} \, [-{1\over 4}F_{\mu\nu}^{a} F^{\mu\nu\ a} +
\bar q( (i\partial_\mu - g A_\mu)\gamma_\mu) - m)q] .\eeq
\noindent
Let us write
\beq q =\left(\matrix{\eta\cr\chi\cr}\right) ,\eeq
\noindent
and choose the gauge $A_- = 0$. Only $\chi$ is
a true degree of freedom. In fact, we can use $\eta$ and $A_+$
equations of motion
\beq \eqalign{i\partial_-\eta &= {m\over 2}\chi  \cr
\partial_-^2 A_+^a &= -{g\over 2} :\chi^\dagger T^a \chi:\, ,\cr}\eeq
\noindent
to write $\eta$ and $A_+^a$ in terms of $\chi$
\beq\eqalign{\eta(x) &= -i{m\over 2}\int dy h(x-y) \chi(y)\cr
A_+^a (x) &= {g\over 2} \int dy G(x-y) :\chi^\dagger T^a \chi (y):,\cr}\eeq
\noindent
where
\beq\eqalign{ h(x) &= {1\over 2}{\rm sgn}(x) ,\cr
 G(x) &= -{1\over 2} |x| .\cr}\eeq
\noindent
The hamiltonian (generator of $x^+$ translations) we obtain is
\beq \eqalign{H = -i{m^2\over 4}&\int dx\, dy
h(x-y) \chi^\dagger_m (x) \chi_m (y)\cr
+ {g^2\over4} &\int dx\,dy G(x-y) :\chi^\dagger_m (x) T_{mn}^a \chi_n (x):
\,\,:\chi^\dagger_{p} (y) T_{pq}^a \chi_q (y):\!,\cr}\label{hchi}\eeq
\noindent
where $m, n \dots$ are color indices. The normal order is in relation
to the vacum defined as
\beq\eqalign{\chi (p) |0> &= 0\cr
\chi^\dagger (-p) |0> &= 0\,\, .  \cr}p>0 \, ,\eeq
\noindent
with
\beq \chi (x) = \int {dp\over {2\pi}}\,  e^{-ipx}\,  \chi (p).\eeq
Defining
\beq M(x^+,x^-,y^-) = {1\over N_c}\,: \chi^\dagger_m (x^+,x^-)
\chi_m (x^+,y^-):,\eeq
\noindent
we can write the hamiltonian\(hchi), after
some rearrangement of the interaction term as
\beq\eqalign{H = -i{\mu ^2 N_c\over 4}&\int dx\,dy h(x-y) M(x,y)\cr
-{g^2 N_c^2\over 4}&\int dx\,dy G(x-y) M(x,y) M(y,x) ,\cr}\label{h m}\eeq
\noindent
with
\beq \mu^2 = m^2 - {g^2 N_c\over \pi}.\eeq
\noindent
The bilocal field $M(x,y)$ creates a quark-antiquark pair (a meson) at two
causally connected points so a theory in terms of these objects do not
violate causality. The commutation relation for the fields $M(x,y)$ are
those of the infinite dimensional general linear group with a central
extension
\beq [M(x,y),M(z,u)] = {1\over N_c}[ \delta (y-z)(M(x,u)-\epsilon (x-u))
- \delta (x-u) (M(z,y) - \epsilon (z-y))],\label{com rel m}\eeq
where
\beq \epsilon (x) = {i\over {2\pi}} {\cal P}({1\over x}).\eeq
\noindent
One particular representation of \(com rel m)
corresponds to two dimensional QCD.
In order to have a theory with no negative
norm states and positive energy this representation should be chosen unitary
and of heighest weight. We can choose the representation this way only when
$N_c$ is an integer [8] (from the point of view of meson theory that is the
only
reason this should be so). The hamiltonian \(h m) and
the commutation relations \(com rel m) define
a quantum theory of mesons equivalent to QCD, but written entirely in terms
of color singlets. It is also Lorentz invariant with the transformation rules
\beq \eqalign{x^+\rightarrow& \lambda x^+\cr
x^-\rightarrow \lambda^{-1} x^-\cr
M(x,y)\rightarrow \lambda M(x,y)\, ,\cr}\eeq
with $\lambda$ related to the boost velocity $v$ by $\lambda = (1-v)(1-v^2)
^{-1/2}$.
Notice that the factor $1/N_c$ plays the
role of $\hbar$. Consequently the classical version of the theory defined
above corresponds to QCD at $N_c\rightarrow \infty$.
The classical equation of motion for $M(x,y)$ is
\beq \eqalign{\partial_+ M(x,y) = {\mu^2\over4}
\int dz &M(x,z)h(z-y)-h(x-z)M(z,y)\cr
-i{g^2N_c\over 2}\int dz\,& G(z-x)M(x,z)(M(z,y)-\epsilon (z-y))\cr
-&G(z-y)M(z,y)(M(x,z)-\epsilon (x-z)) \, .\cr}\label{eq mot m}\eeq
The vacuum expectation value of $M(x,y)$ in QCD is zero. If we expand
\(eq mot m) around $M(x,y)=0$ and keep up to linear terms we have, in
momentum space
\beq \eqalign{\partial_+ M(p,q)
=& {i\mu^2\over 4}( {1\over q} - {1\over p}) M(p,q)\cr
-&{ig^2 N_c\over 4}\int dk\,\,G(k)( {\rm sgn}(p)\,M(p-k,q-k)
- {\rm sgn}(q)\,M(p+k,q+k)).\cr}
\label{meson spec1}\eeq
Defining
\beq\eqalign{P_-\equiv & p-q\cr
\xi\equiv &{p\over P_-}\cr
\eta(p/P_-)\equiv & M(p,q)\cr}\eeq
and remembering that the
Fourier transform of $G(x)$
is a distribution defined by
\beq \int dp\,  G(p) f(p) = \int dp\, {1\over p^2}(f(p) - f(0)) .\eeq
we have
\beq\eqalign{m^2_{\rm meson}\eta(\xi) &= -(m^2 - {N_c g^2\over \pi})
({1\over \xi} + {1\over 1-\xi}) \eta(\xi)\cr
&-ig^2N_c\int d\xi' ({\rm sgn}(\xi)-{\rm sgn}(\xi - 1)) {1\over (\xi - \xi')^2}
(\eta(\xi')-\eta(\xi))\, .\cr}\eeq
This is just the equation derived by \'  tHooft [9] summing the "rainbow"
diagrams whose solution gives the meson spectrum (up to order of $1/N_c$).
As shown by \'  tHooft there are no scattering states, implying confinement.

The baryon number can be writen as
\beq B = \int dx M(x,x).\label{bar numb}\eeq
Thus the nucleon should be identified with the lowest energy static solution
of \(eq mot m) such that $B$ defined by \(bar numb) is equal to $1$.
Due to the complexity of \(eq mot m) we will use numerical analysis and the
ansatz
\beq M(x,y) = {1\over N_c} \Psi^*(x) \Psi (y)
+ \epsilon(x-y)\, ,\label{ansatz}\eeq
with
\beq\int dx |\Psi (x)|^2 = 1 ,\eeq
\noindent
where $\Psi$ is a classical bosonic variable. With this
{\it ansatz} the equation of motion \(eq mot m) reduces to
an equation for $\Psi$
\beq \partial_+\Psi (x) = {m^2\over 4}\int dz h(z-x) \Psi (z)
+ i{g^2\over 2} \int dz G(z-x) |\Psi (z)|^2 \Psi (x)\label{eq mot psi x}\eeq
\noindent
Note we have now $m$ instead of the previous renormalized mass $\mu$.
A static $M(x,y)$ does not imply a static $\Psi(x)$: $\Psi(x)$ can change
in time by a phase
\beq \partial_0 M(x,y)=0\,\,\,\rightarrow\,\,\,
\Psi(x)=
\psi(x^+ - x^-) e^{-i\,f(x^+ + x^-)}.\eeq
This phase can be chosen arbitrarily, without change
of physics. We will choose a phase growing linearly with time
$f(x^+ + x^-)={\cal E}(x^+ + x^-)/2$.
Combining with \(eq mot psi x) we arrive at
\beq \partial_-\psi (x)+ {m^2\over 4}\int dz h(z-x) \psi (z)
+ i{g^2\over 2} \int dz G(z-x) |\psi (z)|^2 \psi (x)=-i{\cal E}\psi(x).
\label{eq mot psi x2}\eeq
\noindent
This is the Hartree equation for relativistic particles interacting through
a Coulomb potential in the light cone variables. This can be seen more
easily if we write \(eq mot psi x2) in momentum space. It will be
more useful though to obtain \(eq mot psi x2) from a variational principle
first. In fact \(eq mot psi x2) is equivalent to minimizing
\beq\eqalign {E=&\int dx \,i\psi^*(x) \partial_-\psi(x)
-i{m^2\over 4}\int dx\,dy \,\,h(x-y)
\psi^*(x)\psi(y)\cr
&-{g^2\over 4}\int dx\,dy\, G(x-y)\,|\psi(x)|^2
|\psi(y)|^2\cr}\label{energy x}\eeq
under the constraint
\beq\int dx\, |\psi(x)|^2 = 1.\label{const}\eeq
\noindent
Notice that the funcional $E$ is just the energy $P_+ + P_-$, $P_+$ being
the hamiltonian $H$ writen in terms of $\psi$. In momentum space $E$ is given
by
\beq\eqalign{E = &\int {dp\over {2\pi}}(p+{m^2\over {4p}})|\psi(p)|^2\cr
-{g^2\over 4}&\int {dp\,dq\,dk\over {8\pi^3}}
\,G(p)\, \psi^*(k)\psi^*(q)\psi(k+p)\psi(q-p) .\cr}
\label{energy p}\eeq
The symmetry $\psi(x)\rightarrow\psi^*(-x)$ of \(eq mot psi x2) implies that
$\psi(p)$ is real for the ground state. Also, in order to
have a positive definite energy
$\psi(p)$ should vanish for negative $p$.
This way we have
\beq\eqalign{E = &\int {dp\over {2\pi}}(p+{m^2\over {4p}})|\psi(p)|^2\cr
-{g^2\over 4}&\int {dp\,dq\,dk\over 8\pi^3}\,
{1\over p^2}\, [\psi(k)\psi(q)\psi(k+p)\psi(q-p)
-\psi(k)^2 \psi(q)^2] .\cr}\label{energy p}\eeq
A discrete version of the energy above was
minimized numerically using the method of the gradient [10],
this means, starting from some arbitrary initial configuration the code changes
it following the negative of the gradient of the discrete energy.
The constraint \(const) was imposed in two steps:
1) projecting out the component
of the gradient normal to the constraint and
2) normalizing the state after
each "time" step. Some examples of wave functions for diverse values of
the parameter $m/g$ are shown in figure 1. They are concentrated around a
mean value of $p_-$ and are smooth, which implies that in
position space $\psi$ vanishes at infinity faster than any polynomial.
Notice that the two terms in $E$ have competing effects.
The kinetic term is minimized with a $\delta$ function around $p=m/2$. The
potential energy though favors a more spread out wave function.
In fact we can see
that the larger the value $g$ for fixed $m$, the broader the function is.
In the weak coupling, non relativistic limit $m/g\rightarrow\infty$, the
peak of the wave function is roughly around $p=m/2$. For the strong coupling
region though, the peak is at larger values of $p$, reaching a value $\sim g/4
$
at $m=0$. At this point the mass of the baryon is different from zero,
signaling that the chiral limit does not correspond to $m=0$ (remember that $m$
is the bare mass of the quark). This is  disturbing and is probably
related to the fact that in \(ansatz) the limit $\Psi(x) = 0$ gives
$M(x,y) \not= 0$. The ground state of the baryon number one sector is not
only $N_c$ quarks on top of the vacum, but contains distortions even far away
from the baryon. Notice it is the $\epsilon(x-y)$ term in \(ansatz) that
produces the mass renormalization back from $\mu$ to $m$.
For negative values of $m^2$ the energy is still
positive as long as $m^2$ is larger than some critical value $\sim -2g^2/3$.
Beyond
this point $E$ is not bounded from below. This behavior of the baryon mass
as function of $m/g$ is shown in figure 2. For large values of $m/g$ the mass
of the baryon (divided by $N_c$, the number of quarks in the baryon)
approaches the quark mass, as it should be for a non
relativistic system. It is a little larger because in two dimensions
the binding energy is positive.
   We also solved \(eq mot psi x) linearised around the wave function
$\psi_0$ of
the (ground state) baryon. Now $\psi (p)$ does not need to be real
(because it describes excited states) and we have
one equation for the real part,
 another for the imaginary part. So, if
\beq \psi(p)= \psi_0(p) + \sigma_R(p) + i\sigma_I(p)\, ,\eeq
then we have
\beq\eqalign{(p+{m^2\over 4p})\sigma_R(p) -{g^2\over 2}\int {dq \, dk\over
(2\pi)^2}
\, G(k-p)& (2 \psi_0(k+q-p)\psi_0(q)\sigma_R(k)\cr
&+\sigma_R(k+q-p)\psi_0(q)\psi_0(k))\cr
&={\cal E}\sigma_R(p)\cr}\eeq
and
\beq\eqalign{(p+{m^2\over 4p})\sigma_I(p) -{g^2\over 2}\int {dq \,dk\over
(2\pi)^2}
\, G(k-p)& (2 \psi_0(k+q-p)\psi_0(q)\sigma_I(k)\cr
&-\sigma_I(k+q-p)\psi_0(q)\psi_0(k))\cr
&={\cal E}\sigma_I(p)\, .\cr}\eeq
The energies ${\cal E}$ correspond
to the excited states of the baryon. Typical examples of the wave function
of the low lying states are shown in figure 3 and 4. Notice
that there are two states
nearly degenerate with the ground state. The source of this degeneracy is
the presence of two symmetries in \(eq mot psi x2). The first is phase
invariance
\beq \psi(x)\rightarrow e^{i\alpha}\psi(x)\eeq
that implies that $i\psi_0(p)$ will be a solution with the same ground state
energy. The second is translation invariance
\beq \psi(x)\rightarrow \psi(x-x_0)\eeq
that implies that $ip\psi(p)$ is also degenerate with the ground state.
Translation symmetry is broken by the lattice so this degeneracy is
slightly lifted in our numerical results, as can be seen
in figure 3. The numerical procedure
to find the n-th excited state was to minimise the energy in the subspace
orthogonal to all states with lower energy, so instead of $ip\psi_0(p)$ we
have a linear combination of $ip\psi_0(p)$ and $i\psi_0(p)$ orthogonal to
$\psi_0(p)$.

One of us (P.B.) would like to acknowledge partial financial support
from CAPES. This work was supported in part by U.S. Department of Energy
Contract No. DE-AC02-76ER13065.

\bigskip
\bigskip
\vfill
\eject
{\bf REFERENCES}
\bigskip
\noindent
[1]T.H.R.Skyrme, Proc. Roy. Soc. A260 (1961) 127;  Nucl. Phys. 31, 556 (1962);
 J. Math. Phys. 12, 1735 (1971)

\noindent
[2]E.Witten, Nucl. Phys. B223 (1983) 422

\noindent
[3]A.Balachandran, V.P.Nair, S.G.Rajeev and A.Stern, Phys. Lett. 49, 1124
(1982)
;  G.Adkins, C.Nappi and E.Witten, Nucl. Phys B228, 552 (1983)

\noindent
[4]\' t Hooft, Nucl. Phys. B72, 461, (1974)

\noindent
[5]E.Witten, Nucl. Phys. B160 (1979) 57

\noindent
[6]K.Hornbostel, S.Brodsky, H.Pauli, Phys. Rev, 41, 3814 (1990)

\noindent
[7]Y.Frishman and J.Sonnenchein, WIS-92-54-PH, July 1992

\noindent
[8]V.Kac and D.H.Peterson, lectures in Seminaire de Math. Sup., Systemes
dynamiques non lin\'eaires, Les Presses de l\' Universit\'e de  Montreal, 1986

\noindent
[9]G.\' tHooft, Nucl. Phys. B75 (1974) 461

\noindent
[10]W.Press, B.Flannery, S.Teukolsky and W.Vetterling, Numerical Recipes: The
Art of Scientific Computing, Cambridge University Press (1986)
\vfill
\eject
{\bf Figure Captions}

\noindent
Figure 1 : Examples of wave functions of the ground state. $P$ is measured
in arbitrary units and $g=400$.

\noindent
Figure 2 : Energy as a function of $m/g$ for positive and negative values of
$m^2$ (at the left of $0$). Energy is measured in arbitrary units and $g=30$.
Negatives values energy are meaningless because the process of minimisation
never reached an end.

\noindent
Figure 3 : Ground state and two other states degenerate with it.

\noindent
Figure 4 : Three examples of excited states.
\end